\documentstyle[preprint,revtex,eqsecnum]{aps}
\begin{document}
\baselineskip=18pt
\font\foot=cmr8
\font\bfoot=cmbx8
\font\buno=cmbx10 scaled \magstep1
\font\bdue=cmbx10 scaled\magstep2
\font\tex=cmr10
\font\it=cmti10
\font\name=cmssi10
\def\capo{\par\noindent}
\def\saltopagina{\par\vfill\eject}
\def\salto{\par\vskip .5cm}
\def\quasisalto{\par\vskip .4cm}
\def\saltone{\par\vskip 2cm}
\def\saltino{\par\vskip .2cm}
\def\e{\epsilon}
\def\em{\epsilon_\mu}
\def\en{\epsilon_\nu}
\def\emn{\epsilon_{\mu\nu}}
\def\enm{\epsilon_{\nu\mu}}
\def\emm{\epsilon_{\mu\mu}}
\def\emnr{\epsilon_{\mu\nu\rho}}
\def\q{\eqno(}
\def\o{\over}
\def\spc{\bf X}
\def\p{\partial}
\def\a{\alpha}
\def\b{\beta}
\def\c{\gamma}
\def\ii2{\int d^2z}
\def\fr{rightarrow}
\def\t{\theta}
\def\f{\phi}
\def\med{{1\o2}}
\def\SP{\sum_{\vec p}}
\newcommand{\lra}{\longrightarrow}
\newcommand{\ra}{\rightarrow}
\newcommand{\nc}{\newcommand}
\nc{\ba}{\begin{array}}
\nc{\ea}{\end{array}}
\nc{\acom}{\renewcommand{\arraystretch}{0.5}}
\nc{\astr}{\renewcommand{\arraystretch}{1}}
\nc{\sst}{\scriptstyle}
\nc{\ts}{\textstyle}
\nc{\ti}{\tilde}
\nc{\la}{\langle}
\nc{\ran}{\rangle}
\nc{\rmed}{ \frac{1}{\sqrt{2}} }
\nc{\factor}{ \sqrt{ \frac{(j+m)!}{(j-m)! (2j)!} } }
\nc{\xx}{{\bf{x}} }
\nc{\zz}{{\bf{z}} }
\nc{\yy}{{\bf{y} }}
\nc{\medi}{{\tiny \med }}
\begin{title}
FACTORIZATION AND DISCRETE STATES \\
IN $\hat C=1$ SUPERLIOUVILLE THEORY
\end{title}
\author{G. Aldazabal}
\begin{instit}
Centro At\'omico Bariloche, 8400 Bariloche, \\
Comisi\'on Nacional de Energ\'{\i}a At\'omica, \\
Consejo Nacional de Investigaciones Cient\'{\i}ficas y T\'ecnicas, \\
and Instituto Balseiro, Universidad Nacional de Cuyo - Argentina
\end{instit}
\author{M.Bonini}
\begin{instit}
Universit\`a di Parma\\
INFN-Sez.Milano-Gruppo collegato di Parma.
\end{instit}
\author{J.M.Maldacena}
\begin{instit}
Centro At\'omico Bariloche, 8400 Bariloche, \\
Comisi\'on Nacional de Energ\'{\i}a At\'omica, \\
and Instituto Balseiro, Universidad Nacional de Cuyo - Argentina
\end{instit}

\begin{abstract}

We study the discrete state structure of
$\hat c=1$  superconformal matter coupled  to 2-D supergravity.
Factorization properties of scattering amplitudes are used to identify these
states and to construct the corresponding vertex operators. For both
Neveu-Schwarz and Ramond sectors these states are shown to be organized in
 SU(2) multiplets.\\
The algebra generated by the discrete states is computed in the limit of
null cosmological  constant.
\end{abstract}

\section{Introduction}
Matrix models and its continuous counterpart, the Liouville model, have been
 subject of intense investigations during the last years as  representations
of 2-$D$ gravity (see \cite{ginsparg,alvarezg,doker} for general references).
The growing interest in 2-$D$ gravity may be in part explained by the need of
finding a non-perturbative scenario for solving different puzzles posed by
perturbative string theory \cite{gross1}. Gravity in two dimensions is
also a tractable toy model for gravitation and it provides a possible
description of
random surfaces \cite{seiberg,alvarezg}.

The coupling of conformal matter to 2-$D$ gravity displays a very rich and non
trivial structure. \cite{polyakov,witten}. For bosonic matter with central
charge $c=1$, the theory can be thought as a string theory in two dimensions.
Though
a ``first sight" counting might suggest that the tachyonic vacuum is the
only degree of freedom, since there are no transverse modes, a more careful
analysis indicates the existence of a full tower of discrete states.
These are remnants of the massive string states
in higher dimensions. Significant progress has been achieved in the
 understanding of this spectrum in the matrix formulation as well as in the
Liouville approach \cite{gross2,polyakovk}.

In the supersymmetric case the situation is less clear. In particular there
is no satisfactory supermatrix formulation
(see however Refs. \cite{alvarezgm,parisi}). On the other hand, recent
works attempt to understand the structure of the scattering
amplitudes  and the spectrum \cite{difran,dhoker2,abda,ohta} in the
super-Liouville model \cite{diku}.

In this article we concentrate in the study of discrete states for $\hat c =1$
(c=3/2) superconformal matter coupled to two dimensional gravity.
By generalizing the analysis performed in
\cite{sakai} to the supersymmetric case, we first look at the pole structure of
tachyonic scattering amplitudes. From the residues at the poles when two
particles collide
to the same point, we read the vertex operators corresponding to
Neveu-Schwarz discrete states. A more complete analysis is then performed
by studying the different constraints on polarization tensors imposed by the
OPE with the superstress energy tensor. The analysis is further generalized to
consider the Ramond sector as well. This is done in sections 2 and 3.

In section 4 we show, by choosing a ``material" gauge
(i.e.  Liouville sector remains in the
ground state), that the discrete states can be
organized into multiplets of a global SU(2) algebra, as in the
bosonic case \cite{polyakov,tasaka}. The generators of the algebra are
obtained by  supercontour integrals of appropriate supercurrents.

In the last section we study the algebra generated by
Neveu-Schwarz and Ramond discrete states and verify that, after proper
redefinitions of the fields, it reduces to the algebra of the area preserving
diffeomorphisms in two dimensions \cite{witten,bou}.
Two appendices are included. In appendix A we show a computation where
three tachyons collide to the same point. In appendix B we define cocycle
factors needed for a correct definition of OPE's.

\section{Scattering amplitudes and Factorization}

Let us start with a brief review of the computations of
Neveu-Schwarz amplitudes for two dimensional supergravity coupled
to N=1 supermatter ($\hat c=1$). \\
In the conformal gauge, matter and super Liouville
theories can be realized by two free superfields ${\bf X}_0$ and ${\bf X}_1$,
the Liouville and the matter superfield:
\begin{equation}
{\bf X}_0=\phi+\t\psi_0,\qquad{\bf X}_1= x+\t\psi_1.
\label{fields}\end{equation}
The total action can be expressed in terms of the
two component superfield ${\bf X}_\mu=({\bf X}_0,{\bf X}_1)$ as follows
\begin{equation}
S={1\o{4\pi}}\int d^2zd^2\t (D_\t{\bf X}\cdot D_{\bar\t}{\bf X}-
iRQ\cdot{\bf X}+2\mu e^{i\a\cdot{\bf X}}),\end{equation}
where  $D=\p_\t+\t\p_z$ is the superderivative. The values of
 $Q_\mu=(-2i,0)$ and \nobreak $\a_\mu=(i,0)$ are respectively fixed by
requiring the
vanishing of the total central charge and that the exponential term in the
 action has the correct conformal weight.

The two point function of the superfield ${\bf X}_\mu$ is given by
\begin{equation}G(1,2)=<{\bf X}_\mu(z_1,\t_1){\bf X}_\nu(z_2,\t_2)>=
-\eta_{\mu\nu}log\vert z_{12}\vert^2 \label{prop}
\end{equation}
where $z_{12}=z_1-z_2-\t_1\t_2$.

The super energy-momentum tensor is
\begin{equation}T(z,\t)=-{1\o2}D^2{\bf X}_\mu D{\bf X}^\mu-
{i\o2}Q_\mu D^3{\bf X}^\mu
\equiv {1\o2} T_F+\t T_B.\end{equation}
The energy momentum
tensor $T_B$ and the super current $T_F$, expressed in terms of the
component fields $X_\mu=(\phi,x)$ and $\psi_\mu=(\psi_0,\psi_1)$,
read
\begin{equation}
T_B=-{1\o2}\p X_\mu\p X^\mu+{1\o2}\psi_\mu\p\psi^\mu-
{i\o2}Q_\mu\p^2X^\mu,
\label{tb}\end{equation}
\begin{equation}T_F=-\psi_\mu\p X^\mu-iQ_\mu\p\psi^\mu
\label{tf}\end{equation}
and satisfy the N=1 superconformal operator product with $\hat c=1$.

The gravitationally dressed tachyon vertex operator $\Psi(p)_{NS}$ of
momentum $p$ is
\begin{equation}\Psi(p)_{NS}=\int d^2zd^2\t e^{iP\cdot{\bf X}}\end{equation}
where $P$ is the two component momentum $P_\mu=(-i\b(p),p)$. By requiring that
 the conformal dimension of $\Psi_{NS}$    is $({1\o2},{1\o2})$, the ``energy"
$\b(p)$ is determined  to be
\begin{equation}\b^\pm=-1\pm p.
\end{equation}
It corresponds to tachyons with chirality $\pm 1$.
The calculation of the N-point function of these tachyons on a sphere is
given by the path integral
\begin{equation}A_N=<\prod_{i=1}^N\Psi_{NS}(p_i)>=
\int{{\it D}{\bf X}_0{\it D}{\bf X}_1\o V} e^{-S}
\int\prod_{i=1}^N \left(d^2z_id^2\t_i e^{iP\cdot{\bf X}(z_i,\t_i)}\right)
\end{equation}
where $V$ is the volume of the gauge symmetry group.
By performing the standard integration over the zero mode of $\phi$ and $x$
this
amplitude becomes
\begin{equation}A_N=({\mu\o{2\pi}})^s\a^{-1}\Gamma(-s)
\int{{\it D}\tilde{\bf X}_0{\it D}\tilde{\bf X}_1\o V}e^{-\tilde S}
\left(\int dzd\t e^{\a\tilde{\bf X}_0}\right)^s
\int\prod_{i=1}^N\left(d^2z_id^2\t_i e^{\b_i\tilde{\bf X}_0}
e^{ip_i\tilde{\bf X}_1}\right)\end{equation}
where
$\tilde S$ is the free action (i.e. $\mu=0$) and
\begin{equation}\sum_{i=1}^N (P_i+s\a+ Q)_\mu=0,
\label{constr}\end{equation}
giving the energy momentum conservation.
Notice that, in order to have a non zero result,
$N+s$ must be even since the tachyon vertex is a fermion operator on the
world-sheet.
This amplitude has been computed in Ref.\cite{difran,abda}  where the two
dimensional integrals have been explicitly evaluated.
 Like in the bosonic case, the amplitude
factorizes in N-external leg factors and exhibits leg poles when the value
of the external tachyon momenta is an integer: $p_i=r+1,\; r=0,1,..$.
Here we study the factorization of the N-tachyon amplitude and we identify
these legs poles as corresponding to higher level states which are present
in the spectrum for discrete value of the momenta.

We consider therefore the following correlation function
\begin{eqnarray}A_{N,s}&=&<\int\prod_{i=1}^{N+s}\left(d^2z_id^2\t_i
e^{iP_i\cdot{\bf X}(z_i,\t_i)}\right)>_0\nonumber\\
&=&\int\prod_{i=1}^{N+s} e^{-\sum_{i\ne j}P_i\cdot P_j G(i,j)}
\label{ans}
\end{eqnarray}
where $<\cdots>_0$ means in the free theory, $P_{N+j}=(\a,0),\; j=1,...s$
and $G(i,j)$ is given in (\ref{prop}).
This amplitude presents singularities when two of the tachyons
collide to a same point (we will comment on the pinching of more
tachyons at the end).

Let us consider the kinematical configuration where all tachyons
except one, have the same positive chirality. By considering
the contribution to the amplitude from the
integration region where one tachyon with momentum $p_1$ and
negative chirality collides with an other tachyon of momentum $p_2$
(and positive chirality), we find
\begin{equation}<...>_0=\int \prod_{i=2}^N dz_i d\t_i
\sum_{m=0}{1\o{2(P_1\cdot P_2+m+1)}}({1\o{m!}})^2
 F_m(2,j),\end{equation}
where
\begin{equation}
F_m(2,j) = < :e^{iP_2\cdot{\bf X}}\vert\p_1^m D_1\vert^2\ e^{iP_1\cdot{\bf X}}:
\prod_{j=3}^N e^{iP_j\cdot{\bf X}}>_0\vert_{1=2}
\label{vert}\end{equation}
(double dots indicate
normal ordering and $\partial \bar\partial {\bf X} \equiv 0$).
Therefore the poles in the amplitude are found for
\begin{equation}P_1\cdot P_2+(m+1)=0
\label{pole}
\end{equation}
where $n=2m+1$ ($n=1,3,...$) is the number of superderivatives in the
 corresponding vertex operator. The intermediate state momentum is $P=P_1+P_2$
and satisfies the level $n$ mass shell condition
\begin{equation}
{1\o2}P(P+Q)+{n-1\o2}=0.
\label{condi}\end{equation}
As expected we find only odd level intermediate states (i.e. bosonic)
in this factorization, due to the fermionic nature of the
tachyon vertex.
Since the kinematical constraint (\ref{constr}) fixes the
value of $P_1$
\begin{equation}p_1=-{N+s-2\o2}=-{t+1\o2}, \qquad \b_1={N+s-4\o2}={t-1\o2}
\label{p1}\end{equation}
where $t=N+s-3$ is an odd number $\ge1$, the condition (\ref{pole})
fixes the value of the momentum
of the other tachyon  and we find a pole in the amplitude
at  level $n$ when $p_2$ is given by
\begin{equation}p_2={t+n\o{2t}}, \qquad
\b_2={n-t\o{2t}}.\label{p2}\end{equation}
Therefore the momentum of the intermediate state is
\begin{equation}p={n-t^2 \o{2t}}, \qquad
\b(p)={t^2-2t+n \o{2t}}= -1+\sqrt{({n-t^2\o{2t}})^2+n}.
\label{inter}
\end{equation}
This is precisely the momentum of a state of level
$n$ with positive chirality.

Notice that from the explicit result of the N-tachyon amplitude, we
know that the amplitude presents leg poles when the momenta of the
positive chirality tachyons are $p_i=r+1$, where $r$ is a
non-negative integer. Then from all possible intermediate states,
only those corresponding to $p_2$ integer survive, the others being null
states.
{}From (\ref{p2}),
the number of tachyons N is then related to the level  $n$ by
$n=(2r+1)t$ and the intermediate
state  momentum (\ref{inter}) is an integer.

Let us first consider levels $n=1$ in the N=4 amplitude for $s=0$
(i.e. $t=1$).
The residue of the pole $n=1$ (i.e. $p_2=1$) is given by
$ F_0(2,j)$
and it is reproduced by the vertex operator (we omit barred derivatives in the
following)
\begin{equation}V_1=\int d\t\em D{\bf X}^\mu e^{iP{\bf X}}
\label{v1}\end{equation}
where $\em=iP_{1\mu}=(0,-i)$ and $P=(0,0)$.
In terms of the component fields this vertex reads $V_1^+=-i\p x$ .
Notice that $\em$ satisfies the polarization condition $\em(P+Q)_\mu=0$.

For $t>1$ (i.e. $N>4$ or $s\ne0$)
the value of $p_2$ at the pole, (\ref{p2}), is not an integer
and therefore we expect to find
only null level $1$ intermediate states.
In fact for general $t$ the residue is
reproduced by a vertex (\ref{v1}) with polarization and momentum  given by
\begin{equation}
\em=({t-1\o2},-i{t+1\o2}),
\qquad P=(-i{(t-1)^2\o{2t}},{1-t^2\o{2t}}).
\end{equation}
For $t\ne1$, this vertex is a total derivative
\begin{equation}V_1(t)={t\o{t-1}}\p(e^{{(t-1)^2\o{2t}}\f-i{t^2-1\o{2t}}x}).
\end{equation}

The residue of the pole $n=3$ (i.e. $F_1(2,j)$ in eq.
(\ref{vert}))
is reproduced by the vertex operator
\begin{equation}
V_3=\int d\t (\em DDD{\bf X}^\mu +
\emn DD{\bf X}^\mu D{\bf X}^\nu) e^{iP{\bf X}}
\end{equation}
where $\em=iP_{1_\mu}$, $\emn=-P_{1_\mu}P_{1_\nu}$ and $P=(P_1+P_2)$
which are given by (\ref{p1}) and (\ref{inter}) with $n=3$.
Non null intermediate states are found for $t=1$ and $t=3$ corresponding to
poles in the amplitude at $p_2=2$ and $p_2=1$ respectively. In the first case
the momentum and polarizations are
\begin{mathletters}
\begin{equation}
P=(-i,1),\quad\epsilon_\mu=(0,-i),\quad
\epsilon_{00}=0,\quad\epsilon_{11}=-1,\quad\epsilon_{01}=\epsilon_{10}=0.
\label{v3ia}
\end{equation}
For the other one we find
\begin{equation} P=(-i,-1),\quad
\epsilon_\mu=(1,-2i)\quad
\epsilon_{00}=1,\quad\epsilon_{11}=-4,\quad\epsilon_{01}=\epsilon_{10}=-2i.
\label{v3ib}
\end{equation}
\end{mathletters}

The operators for higher level
intermediate states may be constructed following the same steps.
In order to obtain even level (i.e. fermionic) intermediate
states a more general factorization of the amplitude must be taken into
account.
In fact, fermionic states may appear only when an odd number of tachyons
collides to a same point on the world-sheet.
In appendix A we will compute the contribution to the amplitude from the
integration region where a negative chirality tachyon collides with two
  positive chirality tachyons .
We will find that the intermediate states have always negative
chirality and therefore the residue vanishes (the left side blob of the
amplitude has two states with negative chirality). This is a general feature
of a factorization which involves more than two tachyons and therefore
the fermionic intermediate states decouple in this amplitude.

The scattering of Ramond fields may also be considered.
In this case it is convenient to bosonize the world-sheet fermion
in order to construct the Ramond vertex\cite{friedan}
\begin{equation}
e^{i\sigma} = {1\o\sqrt{2}}(\psi_0 + i\psi_1),\qquad
e^{-i\sigma} = {1\o\sqrt{2}}(\psi_0 - i\psi_1).\end{equation}
We define the propagator
\footnotemark[1]
\footnotetext[1]{The minus sign inside the logarithm is due to
the $\psi$-propagator we are working with \break $<\psi_\mu(z)
\psi_\nu(z')>= -{\delta_{\mu\nu}\o{z-z'}}$.}
$<\sigma(z)\sigma(z')>=-log(-(z-z'))$.
The Ramond  vertex operator in the $-\med$ picture is
\begin{equation}
V_{-{1\o2}}^\alpha=e^{-\med\rho}e^{\med i\alpha\sigma} e^{iP\cdot X}
\label{ram}\end{equation}
where $\alpha=\pm1$, $\rho$ is the bosonized ghost current
and $P=(-i\beta,p)$.
The energy $\beta$ is determined by imposing the mass shell condition
$\med P(P+Q)+\med=0$ and  that this state is annihilated by $G_0$
(this is equivalent to enforce the  Dirac equation). The two requirements
are satisfied if
\begin{equation}
\beta=-1+\alpha p.\label{dir}
\end{equation}
Recall that in the computation of the amplitude
vertices in the``$\med$"-picture should be considered since the total ghost
 charge must add to $-2$.
The amplitude for $N-2M$ NS-states (tachyons) and $2M$ R-states (eq.
(\ref{ram}))
has been computed in Ref.\cite{difran}. Like in the previous case,
the amplitude is nonvanishing only if all the states, except one, have positive
chirality. When the
negative chirality state is a NS-tachyon (R), $N+s$ must be even (odd)
in order to conserve the total fermionic charge.
Besides the leg poles for integer value of the tachyon momentum,
the amplitude has leg poles when the value of the
R state momentum is $p_i=l+\med,\; l=0,1,..$. These poles may be identified
as corresponding to higher level discrete R states which can be seen in
factorization of the amplitude when one NS state and one R state
of opposite chirality collide to the same point.
Since the derivation is completely similar to the NS case we give
only the result. Intermediate states of level $n$ are found
when $n$ is an even integer
(the mass shell condition and momentum are given by (\ref{condi}) and
(\ref{inter}) respectively). When a negative chirality NS tachyon
collides with the R state, the level $n$ intermediate state corresponds
to the R-leg pole $p_2=l+\med$ for $n=2lt$ and has momentum $p=l-{t\o2}$
(notice that $t=N+s-3$ is odd in this case).
When a negative chirality R state
collides with a NS tachyon, we find an intermediate state for $n=(2r+1)t$
with a momentum $p={2r+1-t\o2}$, corresponding to a NS leg pole at  $p_1=r+1$
(in this case $t$ is even).
Therefore the momentum of the intermediate state is always half-integer.

For example the $n=0$ level state is found only in the first case and
corresponds to the R-leg pole at $p_2=\med$. Its vertex operator is given by
(\ref{ram}) with $\alpha=-1$ and $P=(i{1\o2}, -{1\o2})$.
At level $n=2$ there are two states of momentum $P=(-{i\o2},\pm{1\o2})$
corresponding to the Ramond leg pole $p_2={3\o2}$ and to the tachyon leg
pole $p_1=1$  respectively. The corresponding vertex operators are
\begin{equation}
V_2^+(p=\pm\med)=\{e^{\pm i{3\o2}\sigma}\mp
i(\partial\sigma+\partial x)e^{\mp i{1\o2}\sigma}\}e^{\pm i{x\o2}+{\phi\o2}}
\label{ram2i}
\end{equation}
where $\pm$ is the chirality of the Ramond state (\ref{ram}).

\section{Higher level operators}

In order to gain a better comprehension of these intermediate states it is
useful to look at the OPE of the superstress energy tensor with vertices of
general polarizations. The requirement of conformal invariance imposes
certain conditions on these polarizations. In two dimensions the situation is
rather
peculiar since the number of constraints plus
possible gauge symmetries equals the number of components of the polarization
tensors. Thus, in principle, no degrees of freedom are left. This is true
except for
some particular values of  momentum for which either the constraints are
 relaxed or the gauge transformations become linearly dependent, therefore
leaving space for new states which are responsible for the singularities
of the N-tachyon amplitude.

The general form of the vertex operator for a state of the level $n$
is
\begin{equation}
\int d^2\theta V_R V_L e^{iP{\bf X}}
\end{equation}
where $V_L$ is given by a sum of all possible terms of the form
\begin{equation}
\epsilon_{\mu_1 ...\mu_s}\prod^s_{j=1} D^{n_j}{\bf X}^{\mu_j}
\quad \hbox{such that}\quad
\sum^s_{j=1} n_j =n, \quad n=1,2,3,...
\end{equation}
(in the following we shall consider only the left part of the vertex).
The vertex has conformal dimension $({1\o2},{1\o2})$  if (\ref{condi}) is
satisfied.
The two solutions
\begin{equation}\b=-1\pm\sqrt{p^2+n}
\end{equation}
correspond to ($+$) and ($-$) states. Moreover, the
polarization tensors $\epsilon_{\mu_1 ...\mu_s}$
must satisfy constraints coming from the requirement that
the vertex operator is a primary superfield.
In the following we will find these relations and
the explicit solutions for the $n=1$, $n=2$ and $n=3$ case.

\vskip 20pt

At level $n=1$ the general form of the vertex operator is
\begin{equation}
V_1=\int d\theta\em D{\bf X}^\mu e^{iP{\bf X}}
\end{equation}
and, in this case, eq.(\ref{condi}) reads $P(P+Q)=0$.
The polarization tensor $\em$ must satisfy
\begin{equation}\em(P+Q)_\mu=0.\label{pol1}\end{equation}
This condition is solved by
 $\em=aP_\mu$. In the operator language this solution corresponds to the state
\begin{equation}W_1=G_{-1/2}\vert p>\end{equation}
which is null and therefore decouples
(as usual we  call $G_r$ and $L_m$
the Fourier modes of the supercurrent $T_F$ and $T_B$ respectively).
Indeed this state corresponds
to the gauge symmetry $\em\rightarrow\em+P_\mu$.
Then for a generic value of the momenta
there is no physical degree of freedom at this level.
There are two cases where this reasoning is not valid. The gauge symmetry
does not exist when $P=0$ or the constraint relaxes when $P=-Q$.
In correspondence of these two values of the momentum we have two
physical states
\begin{equation}
V_1^+=\displaystyle\lim_{p\to0}{V_1\over p}=\int d\theta
D{\bf X}_1=\p x
\end{equation}
and
\begin{equation}
V_1^-=\int d\theta D{\bf X}_1e^{-2{\bf X}_0}=
(\p x+2\psi_0\psi_1)e^{-2\phi}.
\end{equation}
The first one is exactly the vertex
(\ref{v1}). The second one
cannot be obtained in the factorization of the N-tachyon amplitude
considered previously (i.e. with one negative chirality tachyon and
the remaining ones of positive chirality)
since they decouple in this amplitude.

At level $n=2$ the vertex operator is given by
\begin{equation}
V_2=(\epsilon_\mu DD{\bf X}^\mu +
\epsilon_{\mu\nu}D{\bf X}^\mu D{\bf X}^\nu)e^{iP{\bf X}}
\end{equation}
where $P(P+Q)+1=0$, $\epsilon_{\mu\nu}=-\epsilon_{\nu\mu}$ and
\begin{equation}
\epsilon_\nu - 2i(P+Q)_\mu \epsilon_{\mu\nu}=0 .
\end{equation}
The only solution is
\begin{eqnarray}
\epsilon_\mu&=&(-4+P.Q)P_\mu+Q_\mu \nonumber\\
\epsilon_{\mu\nu}&=&{i\o2}(P_\mu\en-P_\nu\em).
\end{eqnarray}
These polarizations
correspond in the operator language to the state
\begin{equation}
W_2=G_{-1/2}(\epsilon\cdot\psi e^{iPX})
\end{equation}
which is null.
It is easy to check that neither the gauge symmetry  degenerates nor
the constraints relax for physical states. This result is consistent with
the Kac-determinant argument \cite{ohta}.

At level $n=3$ we have
\begin{equation}
V_3=(\epsilon_\mu DDD{\bf X}^\mu+
\epsilon_{\mu\nu}DD{\bf X}^\mu D{\bf X}^\nu)e^{iP{\bf X}}
\label{v3}
\end{equation}
where $P(P+Q)+2=0$.
In general it is possible to include a term of the form
$\epsilon_{\mu\nu\rho}D{\bf X}^\mu D{\bf X}^\nu D{\bf X}^\rho$.
However, due to the fact that $\epsilon_{\mu\nu\rho}$ must be completely
antisymmetric this term is not present in two  dimensions.
The following constraints must be satisfied
\begin{equation}
\epsilon_{\mu\mu}+i\epsilon_\mu (P+2Q)^\mu =0,\label{pmm}\end{equation}
\begin{equation}\epsilon_\mu -i\epsilon_{\mu\nu}(P+Q)_\nu =0,
\label{pn}\end{equation}
\begin{equation}\epsilon_{\nu\mu}=\epsilon_{\mu\nu}.
\label{pmn}\end{equation}
The number of degrees of freedom is therefore reduced  to two by these
equations. This is equal to the number of gauge symmetries generated
by the two independent null states
\begin{mathletters}
\begin{equation}
W_3^{(a)}=(G_{-3/2}+2G_{-1/2}L_{-1})\vert p>
\label{v3a}
\end{equation}
\begin{equation}
W_3^{(b)}=G_{-1/2}(\epsilon\cdot\partial X e^{iPX})
+L_{-1}(\epsilon\cdot\psi e^{iPX})\label{v3b}
\end{equation}
\end{mathletters}
where
\begin{equation}
\epsilon_\mu (P+Q)_\mu=0.\label{pb}
\end{equation}
The vertex operators $V_3^{(a)}$ and $V_3^{(b)}$
corresponding to these states are given by
(\ref{v3}) with the following polarizations
\begin{mathletters}
\begin{equation}
\epsilon_\mu^{(a)}=i({3\o2}P_\mu -{1\o2}Q_\mu),\;
\quad\epsilon_{\mu\nu}^{(a)}= -{1\over2}\eta_{\mu\nu}
-P_\mu P_\nu
\label{eam}\end{equation}
\begin{equation}\epsilon_\mu^{(b)}=i\{(-4+P\cdot Q)P_\mu+2Q_\mu\},\;\quad
\epsilon_{\mu\nu}^{(b)}=
{i\over2}(P_\mu\epsilon_\nu^{(b)}+P_\nu\epsilon_\mu^{(b)}).
\label{ebm}\end{equation}
\end{mathletters}
It is easy to check, by using (\ref{condi}), that the polarizations
(\ref{eam}), and (\ref{ebm}) satisfy
(\ref{pmm}-\ref{pmn}).

Therefore the most general vertex operator for a state of the level $n=3$
is null and it is given by a linear combination of $V_3^{(a)}$ and $V_3^{(b)}$.
However for the particular value of  momentum
$P=(-i,\pm 1)$, the two operators (\ref{v3a}) and (\ref{v3b})
become linearly dependent and it is possible to find a combination
of them with zero polarization tensors, therefore trivially satisfying
eqs. (\ref{pmm})-(\ref{pmn}).
As in the $n=1$ level, a non null
vertex operator can be found by taking a particular limit to
this value of the momentum
\begin{equation}V_3^+(p=\pm1)=\displaystyle\lim_{p\to\pm 1}
{V_3^{(a)}+{1\o 4}V_3^{(b)}\over{p\mp 1}}.
\label{v3pm}
\end{equation}
By using eqs. (\ref{eam}) and (\ref{ebm}) the polarization tensors are
\begin{equation}
\epsilon_\mu=(\mp{1\o4},-{i\o4}),\quad\epsilon_{00}=\mp {1\o4},
\quad\epsilon_{11}=\pm {5\o4},\quad \epsilon_{10}=\epsilon_{01}=-{i\o2}.
\label{pd3}\end{equation}
They satisfy the polarization equations (\ref{pmm}-\ref{pmn}).
The two vertex operators (\ref{v3pm}) coincide with those found in
the factorization of the 4-tachyon amplitude
up to null operators.
In fact, by using (\ref{eam}-\ref{ebm}) and (\ref{v3ia}-\ref{v3ib}) we find
${1\o2}V_3=V_3^+ -V_3^{(a)}-{3\o2}V_3^{(b)}$.
Notice that by performing the gauge transformation which corresponds to the
addition of the null state $\mp{1\o8}V_3^{(b)}$  to $V_3^+(p=\pm1)$,
it is possible to change to a state possessing only matter excitations
(the so called material gauge). This state has
$\epsilon_1=i,\;\epsilon_{11}=\pm1$
and all the other components equal to zero.

Other non null states may be found  when
the constraints eqs.(\ref{pmm})-(\ref{pmn}) relax. This happens when
$P=(3i,\pm1)$. For each value of $P$ there is a ($-$) state
with a polarization given by
\begin{equation}
\e_\mu=(0,0),\qquad\e_{00}=-\e_{11}=\pm i\e_{01}=\pm i\e_{10}=1.
\end{equation}
Also in this case it is possible to make a gauge  transformation to the
material gauge by adding the null state ${1\o8}(V_3^{(a)}-{11\o4}V_3^{(b)})$.

The above analysis can be repeated for the Ramond sector of the theory.
Again physical states are allowed only for
some particular value of the momenta.
In the following we explicitly work out the polarization conditions and
their solutions for the level $n=2$  and compare the result with the
states found previously in the factorization (eq.(\ref{ram2i})).
The general form of the vertex operator is
\begin{equation}
V_2=e^{-{\rho\o2}}\{(\e_+^\a\p X^-+\e_-^\a\p X^+ +
\omega^\a\p\sigma)e^{{i\o2}\a\sigma}+v^\a e^{i{3\o2}\a\sigma}\}e^{iP\cdot X}
\label{v2r}\end{equation}
  where $\a=\pm1$ (we have introduced the complex notation
$X^\pm={1\o \sqrt2}(\phi\pm ix)$). The mass shell condition for this
state is given by (\ref{condi}) with $n=2$. The polarization $\e_\pm^\a$,
$\omega^\a$ and $v^\a$ must satisfy certain conditions coming from the
requirement of superconformal invariance ($G_0$ and $G_1$ must annihilate the
state). By using (\ref{tf}) we find
\begin{mathletters}
\begin{eqnarray}
v^+&=&-i\e_-^-(P_-+\med Q_-),\qquad
\omega^-=\e_+^-(P_-+\med Q_-),\nonumber\\
\e_-^-&=&(P_-+{3\o2}Q_-)\omega^-+i(P_++{3\o2}Q_+)v^+
\label{cos1}\end{eqnarray}
and
\begin{eqnarray}v^-&=&-i\e_+^+(P_++\med Q_+),\qquad
\omega^+=-\e_-^+(P_++\med Q_+),\nonumber\\
\e_+^+&=&i(P_-+{3\o2}Q_-)v^--(P_++{3\o2}Q_+)\omega^+
\label{cos2}\end{eqnarray}
\end{mathletters}
where $P_\pm={1\o\sqrt2}(P_0\pm iP_1)={-i\o\sqrt2}(\beta\mp p)$.
The solutions are
\begin{mathletters}
\begin{eqnarray}
\e_+^-&=&2iP_+,\qquad\e_-^-=i(P_--\med Q_-),\nonumber\\
v^+&=&(P_-+\med Q_-)(P_--\med Q_-),
\qquad\omega^-=-i(1+Q_+P_-)
\label{gau1}\end{eqnarray}
and
\begin{eqnarray}
\e_-^+&=&2iP_-,\qquad\e_+^+=i(P_+-\med Q_+),\nonumber\\
v^-&=&(P_++\med Q_+)(P_+-\med Q_+),\qquad\omega^+=i(1+Q_-P_+).
\label{gau2}\end{eqnarray}
\end{mathletters}
They correspond to the null states $G_0G_{-1}\vert p,->$ and
$G_0G_{-1}\vert p,+>$ respectively,
where  $\vert p,\a>$ is the state (\ref{ram}) with $P(P+Q)=-1$.
However there is a particular value of the momenta for which the
gauge transformation degenerates. In the first case (eq.(\ref{gau1}))
this happens for $P_+=0$ and $P_-={1\o2}Q_-$, i.e. $P=(-{i\o2},{1\o2})$,
in the second (eq.(\ref{gau2})) for $P_-=0$ and $P_+={1\o2}Q_+$, i.e.
$P=(-{i\o2},-{1\o2})$, leading to  ($+$) states (\ref{ram2i}).

Other non null states are found when the constraints
(\ref{cos1}) and (\ref{cos2}) relax. This happens when $P_-Q_+=3$
and $P_+Q_-=3$ (i.e $P=(i{5\o2},\mp{1\o2})$), respectively.
For these values of the momentum there are two physical states.
Up to a gauge transformation their vertex  is
\begin{equation}
V_2^-(p=\mp{1\o2})=e^{-{\rho\o2}}\{i(\p x-\p\sigma)e^{\mp{i\o2}\sigma}
\pm e^{\pm i{3\o2}\sigma}\}e^{{\mp i\o2}x-{5\o2}\phi}.
\end{equation}

\section{SU(2) current algebra and discrete states}

In the bosonic $c=1$ theory, further insight
on the structure of  the discrete states can be gained
by  realizing  that they, together with some tachyon states, fit
into multiplets of an SU(2) algebra.
The highest weight vectors of the SU(2) multiplets are  tachyon states
with momentum $p = \sqrt{2}j ;~j = 1/2,~1,~3/2, \cdots $.
and  discrete states are obtained applying the lowering operator of the SU(2)
algebra.
All  primary states of the $c=1$
theory at the SU(2) compactification radius can be obtained with this
procedure.
We will now show how to make a similar analysis for the N=1 case. In the
bosonic theory the generators have the key property  that
they do not change the conformal properties
of the state on which they act. This is ensured because they are
contour integrals of  conformal weight 1 currents.
 It is easy to show
that the supercontour integral of a
superconformal superfield $j_{1/2}(z,\theta ) $ of weight $h=1/2$ commutes with
the super Virasoro algebra. In fact
\begin{eqnarray}
 [ \oint \frac{dz'}{2 \pi i} d\theta'
 \epsilon (z',\theta') T_{z\theta}(z',\theta')~
 &,&~ \oint \frac{dz}{2 \pi i} d\theta j_{h}(z,\theta) ]\nonumber\\
&=&\oint \frac{dz}{2 \pi i} d\theta [\epsilon \partial + \med (D\epsilon) D +
h (\partial \epsilon) ] j_h (z,\theta)
\end{eqnarray}
and the r.h.s is the contour integral of a total derivative if $h=1/2$.
The SU(2) generators are given by the following contour integrals
\begin{equation}
H_\pm =\pm \sqrt{2}  \oint \frac{dz}{2 \pi i} d\theta
e^{\pm i{\bf X}_1(z,\theta)}
{}~~~~~~~~~~~
H_0 = \oint \frac{dz}{2 \pi i} d \theta iD{\bf X}_1(z,\theta).
 \label{gen}
 \end{equation}
In the  Neveu-Schwarz  sector the primary fields for the discrete states
can be constructed using this algebra as follows.
The highest weight vector is the matter part of a tachyon
\begin{equation}
\phi_{jj}(z,\theta) = e^{i j {\bf X}_1(z,\theta) }
 \label{} \end{equation}
where the momentum must be integer so that the action of $H_\pm$ is well
defined. The states are obtained as
\begin{equation}
\phi_{jm}(z,\theta) =k(j,m) (H_-)^{j-m} \phi_{jj}~~~~~~~k(j,m)=\factor
 \label{} \end{equation}
where $j=0,1...$ and  $m=-j,-j+1,..j$. They
have conformal weight $j^2/2$ and momentum $p= m $. These are precisely the
weights and momenta for the states in the NS sector
prescribed by the Kac formula argument of ref. \cite{ohta},
By adding the Liouville field part with the two possible dressings
we obtain
\begin{equation}
V_{jm}^{\pm} = \int d\theta \Phi_{jm}^{\pm}
\label{estns} \end{equation}
where
\begin{equation}
\Phi_{jm}^\pm = \phi_{jm} e^{(-1 \pm j) {\bf X}_0}.
\end{equation}
According to our previous convention, the level of these states is
$n=j^2-m^2$.

In the Ramond case it is convenient to include the Liouville field
from the beginning since the highest weight vectors of the SU(2) algebra
are the Ramond ground states with half-integer momentum
\footnotemark[1]
\footnotetext[1]{We use the vertex in the $-{1\o2}$-picture. As noted in
\cite{friedan} there are infinitely many operators with different ghost number
for each physical state.}
\begin{equation}
R_{jj}^\pm =e^{-\med\rho(z) }
e^{\pm \med i \sigma(z)} e^{ij x(z)} e^{(-1 \pm j) \phi}.
\label{ramd} \end{equation}
Notice that if $j$ is half integer
the action of $ H_\pm $ is well defined as can be seen by
using the bosonized form of these operators
\begin{equation}
H_{\pm} = \oint \frac{dz}{2 \pi i} \left( e^{i \sigma(z)} - e^{-i \sigma(z)}
\right) e^{\pm i x(z) }.
\label{hmenr} \end{equation}
The rest of the multiplet is obtained acting with $H_-$
\begin{equation}
R^{\pm}_{jm} = k(j,m) (H_-)^{j-m} R^\pm_{jj}.
 \label{estr} \end{equation}
It is easy to see, analysing the weight and momentum of the matter part, that
all the states predicted by the Kac formula \cite{ohta} for the Ramond case
are obtained in this way.

\section{Algebra of discrete states}

The description of discrete states as multiplets of SU(2) is very useful
since the study of the interaction between these modes is greatly simplified
demanding  SU(2) covariance \cite{polyakovk,witten}.
Their
interaction  is given by the quadratic piece in the
$\beta$-function which in turn depends on the coefficient of term
$1/z$ of their OPE
\begin{equation} {\cal O}^{\pm}_{j_{1}m_{1}}(z)
{\cal O}^{\pm}_{j_{2}m_{2}}(z')
\sim \frac{1}{z-z'}
F_{j_{1}m_{1},j_{2}m_{2}}^{j_{3}m_{3}}
{\cal O}^{\pm}_{j_{3}m_{3}} (z')
 \label{ope} \end{equation}
where ${\cal O}$ denotes a NS or R state (\ref{estns},\ref{estr}). We first
take the
plus sign on both fields on the l.h.s., other cases will be analysed
later on.
Conservation of matter momentum implies that
 $m_{3}=
m_{1} +m_{2} $ and conservation of Liouville momentum (valid on the bulk) says
that we have also the plus sign on the r.h.s. of (\ref{ope}) and
that $j_{3}
= j_{1} + j_{2} -1 $.
As the left hand side
of (\ref{ope}) transforms as a product of two SU(2) representations we have
\begin{equation}
 F_{j_{1},m_{1},j_{2},m_{2}}^{j_{1} + j_{2} -1,m_{1}+m_{2}}  =
C_{j_{1},m_{1},j_{2},m_{2}}^{j_{1} + j_{2} -1,m_{1}+m_{2}} g(j_{1},j_{2})
 \label{defdef} \end{equation}
where $C_{j_{1},m_{1},j_{2},m_{2}}^{j_{1} + j_{2} -1,m_{1}+m_{2}}$ is the
Clebsh-Gordan coefficient
for the product of two SU(2) representations, which for the above values of
$j_3$ and $m_3$ reads
\begin{equation}
 C_{j_{1},m_{1},j_{2},m_{2}}^{j_{3},m_3} =
\frac{N(j_{3},m_{3})}{ N(j_{1},m_{1})  N(j_{2},m_{2}) }
\frac{j_{2} m_{1} - j_{1} m_{2} }{\sqrt{j_3(j_3 +1)} }
\label{} \end{equation}
where
\begin{equation}
 N(j,m) = \left[\frac{ (j+m)! (j-m)!}{(2j-1) !} \right] ^{1/2}.
\label{} \end{equation}
The value of $g(j_1,j_2) $ can be found performing explicitly
the OPE for specific values of $m_{1}$ and $ m_2$. We choose
 $m_{1} = j_1 -1, ~m_2 = j_2 $.
We consider separately the three possibilities, i.e. two NS states, one NS
and one R and finally two R states in (\ref{ope}), since
the corresponding functions $g(j_1,j_2)$ are different.
Moreover it is necessary to include cocycle factors to ensure proper
commutation
relations. Our cocycles are constructed in appendix B.
Let us start with two NS states
$$
V^+_{j_1,j_1-1}(z) V^+_{j_2,j_2}(z')
 = \frac{1}{\sqrt{ 2 j_1}} H_{-}(
\int d\theta \Phi^+_{j_1,j_1}) \int d \theta' \Phi^+_{j_2,j_2}
$$
$$ =\frac{-1}{\sqrt{j_1}} \int d \theta  d \theta' \oint
\frac{dud\theta_u}{2 \pi i}
:e^{-i x (z+u,\theta_u)} ::e^{i j_1 x(z,\theta)+(-1+j_1)\phi(z,\theta)}:
:e^{i j_2 x(z',\theta')+(-1 + j_2)\phi(z',\theta')}:
$$
\begin{eqnarray}
 = \frac{-1}{\sqrt{j_1}}
\int d \theta d \theta'  \oint \frac{du}{2 \pi i} d\theta_u
(u-\theta_u\theta)^{-j_1}
(z+u-z'&-&\theta_u\theta')^{-j_2} (z-z'-\theta\theta')^{j_1+ j_2 -1 }
\nonumber\\
&:& e^{ij_{3}x+(-1 + j_3)\phi}:
\nonumber \end{eqnarray}
No derivatives of the exponential appear because we are interested in
the most divergent term in the OPE.
Performing the $\theta$-integrations and the rescaling $u = (z-z')y $ we obtain
\begin{equation}
\frac{1}{z-z'} \frac{1}{\sqrt{j_1}}  \oint \frac{dy}{2 \pi i}
y^{-j_1}\partial_y (1+y)^{-j_2}  \int d \theta'\Phi_{j_3j_3}(z',\theta)
\end{equation}
Finally computing the contour integral
and considering the appropiate cocycle factor $ (-1)^{j_1 + 1} $ we have
\begin{equation}
 F^{j_1 +j_2 -1,j_1 +j_2 -1}_{j_1,j_1-1,j_2,j_2} =
- \frac{(j_1 + j_2 -1)!}{\sqrt{j_1} (j_1-1)! (j_2 -1)!}.
\end{equation}
Upon comparing  this with (\ref{defdef}) we find $g(j_1,j_2)$ and replacing
this value back in (\ref{defdef}) we find
\begin{equation}
 F_{j_{1},m_{1},j_{2},m_{2}}^{j_{1} + j_{2} -1,m_{1}+m_{2}}  =
 \frac{N(j_3,m_3) (j_3-1)!\sqrt{j_3} }{N(j_1,m_1)(j_1-1)! \sqrt{j_1}
 N(j_2,m_2) (j_2 -1)! \sqrt{ j_2 } }
(j_2 m_1 - j_1 m_2).
\end{equation}

In the OPE of one NS state with one
R state it is convenient to use the bosonized form of the NS vertex
(\ref{estns})
\begin{equation}
V_{jj}^+ =
\rmed \left[ (-1 + 2j)e^{i\sigma} - e^{-i\sigma} \right]
e^{i j x} e^{(-1+j) \phi}.
 \label{} \end{equation}
As in the previous case, we calculate $g(j_1,j_2)$ from the operator product
\begin{equation}
 V^+_{j_1,j_1-1}(z) R^+_{j_2,j_2}(0) = \frac{1}{\sqrt{ 2 j_1}} H_{-}(
V^+_{j_1,j_1})(z) R^+_{j_2,j_2}(0).
\end{equation}
After a calculation similar to the one sketched above, we  obtain
\begin{equation}
 F_{j_{1},m_{1},j_{2},m_{2}}^{j_{1} + j_{2} -1,m_{1}+m_{2}}  =
\frac{ N(j_3,m_3) (j_3 -1/2)! \sqrt{j_3} }{N(j_1,m_1)
(j_1 -1)! \sqrt{j_1}  N(j_2, m_2) (j_2 -1/2)! \sqrt{j_2}} (j_2 m_1 - j_1 m_2).
 \label{} \end{equation}

When the operator product of two R states (\ref{estr}) is considered,
 a NS state with
superghost number $-1$ is found. This operator  is related through picture
changing\footnotemark[1]
\footnotetext[1]{For example the tachyon vertex in the $-1$ picture reads
$V_{jj}^{\pm~(-1)} = e^{ijx}e^{(-1+j)\phi} $
(as explained in \cite{friedan}). }
 to the conventional NS operator of
superghost number zero (\ref{estns}).
Eq.(\ref{ope}) for this case reads
\begin{equation}
R^+_{j_1m_1}(z) R^+_{j_2m_2}(z') \sim \frac{1}{z-z'}
 F_{j_{1},m_{1},j_{2},m_{2}}^{j_{3},m_{3}}  V^{+ (-1)}_{j_3 m_3}.
 \label{operr} \end{equation}
Performing explicitly the
operator product (\ref{operr}) for the case $m_1 =j_1 -1,~m_2 = j_2$
(in order to extract $g(j_1,j_2)$) and considering the cocycle factor
$(-1)^{j_1+\med}$,
 we finally obtain
\begin{equation}
 F_{j_{1},m_{1},j_{2},m_{2}}^{j_{1} + j_{2} -1,m_{1}+m_{2}}  =
   \frac{
N(j_3,m_3)  (j_3 -1)! \sqrt{j_3} }{
N(j_1,m_1) (j_1 -1/2)! \sqrt{j_1}
N(j_2,m_2)  (j_2 -1/2)! \sqrt{j_2} }
\frac{1}{\sqrt{2}} (j_2 m_1 -j_1 m_2).
\end{equation}
After the following redefinitions
\begin{equation}
{\cal O}^+_{jm} \ra \ti{N}_(j,m) {\cal O}^+_{j,m}
{}~~~\mbox{ with}~~~
\ti{N}(j,m) = \left\{ \ba{ll}  N(j,m) (j-1)! \sqrt{j}~&~j \in {\bf Z} \\
  2^{1/4}
 N(j,m) (j-1/2)! \sqrt{j}  ~&~j \in {\bf Z} + 1/2  \ea \right.
\end{equation}
the structure constants are
\begin{equation}
 F_{j_{1},m_{1},j_{2},m_{2}}^{j_{1} + j_{2} -1,m_{1}+m_{2}}  =
(j_2 m_1 -j_1 m_2)
\end{equation}
for all cases.

A similar analysis can be performed for the operator product coefficients of
the
fields ${\cal O}^-$. When we have two minus signs in (\ref{ope}), conservation
of Liouville momentum implies that $j_3 = j_1 + j_2 +1$. As this representation
does not appear in the product $j_1 \otimes j_2$ we conclude that
\begin{equation}
{\cal O}^-(z)
{\cal O}^-(z') \sim  0\times \frac{1}{z-z'}.
\end{equation}
Using similar arguments for the case of ${\cal O}^+ {\cal O}^-$ we see that
\begin{equation}
{\cal O}^+_{j_1m_1}(z) {\cal O}^-_{j_2m_2}(z') \sim \left\{ \ba{lll}
 0 \times \frac{1}{z-z'}  & ~~~\mbox{for}~~~ & j_2 \leq j_1 -1 \\
\frac{1}{z-z'}
F_{j_{1},m_{1},j_{2},m_{2}}^{+-~j_{3},m_{3}}
{\cal O}^-_{j_3 m_3}
& ~~~\mbox{for}~~~ &
j_2 > j_1 -1 \ea  \right.
 \label{opemenos} \end{equation}
where $m_3 = m_1 + m_2,~j_3 = j_2 -j_1 +1 $.
In principle we could
 find these OPE coefficients using SU(2) covariance and
performing explicitly some operator products. We note, however,
 that they can be obtained
by using associativity of the OPE in
the three point function
\begin{equation}
\langle ({\cal O}^{r_1~+}_{j_1m_1} {\cal O}^{r_2~+}_{j_2m_2} )
{\cal O}^{r_3~-}_{j_3~{-m_3}}
 \rangle  = \langle
{\cal O}^{r_1~+}_{j_1m_1} ({\cal O}^{r_2~+}_{j_2m_2}
{\cal O}^{r_3~-}_{j_3~{-m_3}}) \rangle
\nonumber \end{equation}
then
\begin{equation}
F_{j_{1}m_{1},j_{2}m_{2}}^{++~j_{3}m_{3}}
\langle {\cal O}^{r_1+r_2~+}_{j_3m_3}
{\cal O}^{r_3~ -}_{j_3~{-m_3}} \rangle =
F_{j_{2}m_{2},j_{3}~{-m_{3}}}^{+-~j_{1}{-m_{1}}}
\langle {\cal O}^{r_1+}_{j_1m_1}
{\cal O}^{r_2+r_3 -}_{j_1~{-m_1}} \rangle
\nonumber \end{equation}
where $r_i$ is the superghost number ($r_1 +r_2 +r_3 =-2 $) and
$j_3 =j_1 +j_2 -1,~m_3 =m_1 +m_2 $. By deforming contours we find that
\begin{equation}
\langle {\cal O}^{r~+}_{jm}
{\cal O}^{-2-r~ -}_{j~{-m}} \rangle
= (-1)^{j-m}
\langle {\cal O}^{r~+}_{jj}
{\cal O}^{-2-r~ -}_{j~{-j}} \rangle
= (-1)^{j-m} s(j) \frac{1}{(z-z')^2} .
 \label{} \end{equation}
Where $s(j)$ is a factor whose explicit expression
 is unnecessary for our purposes.
Thus after renormalising the operators
\begin{equation}
{\cal O}^{r-}_{jm} \ra \frac{1}{(-1)^{j-m} s(j) \ti{N}(j,m) }
{\cal O}^{r-}_{jm}.
 \label{} \end{equation}
the structure constants are
\begin{equation}
F_{j_{1}m_{1},j_{2}m_{2}}^{+-~j_{3}m_{3}} =
 -(j_3m_1+j_1m_3),
 \label{} \end{equation}

We find that, after renormalizing the operators, the algebra is the same as in
the bosonic case.
The fact that we had three different
cases, NS-NS, NS-R and R-R, and only two possible ways of redefining the
operators is a check of our computations.
Our work also provides an alternative,
and  more explicit, derivation of the result of ref. \cite{bou}.
We have computed
the algebra only for the left sector (or open string), for the closed string
we should join the right one.

A space time interpretation of these discrete higher level states is needed
both in the open an closed string case. In the latter,
the GSO-projection must be included. Actually,
in $d=2$ it is not necessary to make this projection, since there is no
true tachyon, but, once it is done, the theory should become topological, as
it is conjectured in ref. \cite{difran}. It could be interesting to compute
the effective action in this case.

\acknowledgments
We are greatful to C.Nu\~nez for many valuable suggestions. G.A would like
to thank the University of Parma for hospitality and M.B. would like to thank
the Centro At\'omico Bariloche where part of this work was done.
\appendix{}

In this appendix we study the factorization process where one tachyon of
momentum $p_1$
and  positive chirality collides with two tachyons, one of momentum
$p_2$ and positive chirality and an other of momentum $p_3$
and negative chirality, (i.e. the contribution from the region where
$z_2\rightarrow z_1$ and $z_3\rightarrow z_1$ in (\ref{ans})).
\begin{equation}
<...>_0=\int\vert \prod_{i=1}^N dz_i d\t_i\sum_{n,k=0}{1\o{n!k!}}
z_{21}^{-\nu_{12}+n}z_{31}^{-\nu_{13}+k} z_{32}^{-\nu_{32}}
F_{nk}(1,j)\vert^2\label{a3t}\end{equation}
where $\nu_{ij}=-P_i\cdot P_j$, $z_{12}=z_1-z_2-\t_2\t_1$
and similarly for $z_{31}$, $z_{32}$.
$F_{nk}(1,j)$ is (the left part of)
\begin{equation}
F_{nk}(1,j)=<:e^{iP_1\cdot{\bf X}}\partial_2^n(1+\t_{21}D_2)
e^{iP_2\cdot{\bf X}}\partial_3^k(1+\t_{31}D_3) e^{iP_3\cdot{\bf X}}:
\prod_{j=4}^N e^{iP_j\cdot{\bf X}}>_0\vert_{2=1,3=1}.
\end{equation}
By performing the rescaling $z_3-z_1=uv$ where $u=z_2-z_1$ and
integrating over $u$, the poles are found for $\nu=l+1$, where
$l=0,1,...$ and $\nu=\nu_{12}+\nu_{13}+\nu_{23}$. The momentum of the
corresponding intermediate states is given by
$P=(-i(p_1+p_2-2+{t-1\o2}),p_1+p_2-{t+1\o2})$. By using the pole
condition
\begin{equation}\nu=2-t-(p_1+p_2)(1-t)=l+1\end{equation}
it is easy to show that this is a momentum of a level $n=2l$ state
with negative chirality.
Therefore the residue at these poles should be zero. We have checked this
for the intermediate tachyon ($l=0$).
In this case the residue is given by
\begin{eqnarray}
A_0=\int&&\vert dz_1 d\t_1dv\prod_{i=4}^N dz_i d\t_i
v^{-\nu_{31}}(v-1)^{-\nu_{23}}
\{-{\nu_{23}\o{v-1}}+{\nu_{31}\o v}\t_1D_2+\nonumber\\
&+&(\nu_{23}+\nu_{31}-1)\t_1D_3\}\vert^2
<:e^{iP_1\cdot{\bf X}}e^{iP_2\cdot{\bf X}}e^{iP_3\cdot{\bf X}}:
\prod_{j=4}^N e^{iP_j\cdot{\bf X}}>_0\vert_{2=1,3=1}.
\end{eqnarray}
The result of the integration over $v$ is zero.
For example the first term gives
\begin{equation}
\nu_{23}\int d^2v \vert v\vert^{-2\nu_{31}}\vert v-1\vert^{-2\nu_{23}-2}=
\end{equation}
\begin{equation}
=\nu_{23}{\Gamma(-\nu_{31}+1)\Gamma(-\nu_{23})\Gamma(\nu_{31}+\nu_{23})
\o{\Gamma(\nu_{31})\Gamma(\nu_{23}+1)\Gamma(-\nu_{23}-\nu_{31}+1)}}=0
\end{equation}
since $\nu_{23}+\nu_{31}=1$ due to energy momentum conservation and the
pole condition ($\nu=1$). The other two terms are zero for the same reason.

\appendix{Cocycle operators}

\renewcommand{\arraystretch}{0.5}
The operators (\ref{estns},\ref{estr}) do not commute inside radial ordered
correlation functions.  The commutation of the exponential
part of them, $e^{r\rho} e^{i\alpha\sigma} e^{imx} e^{\beta\phi}$, gives
rise to a factor $(-1) ^{{\bf x}_1\cdot{\bf x}_2}$ where
${\bf x}=(r,\a,m,\beta)$ and the scalar product is defined with the metric
$(-,+,+,-)$. Notice that ${\bf x}_1\cdot{\bf x}_2$ is an integer since the
 operators are mutually local.
In order to ensure the correct commutation relations cocycle operators
are needed. \cite{goddard}.
It is convenient to express $\xx $ as a linear combination with
integer coefficient of some basis vectors
\begin{equation}
\xx = n_i e_i = {\tiny \frac{1}{2} } (m +\beta +\alpha +r) e_1 + \medi ( m +
\beta -\alpha -r)
e_2 -2\beta e_3 +(-\beta +r) e_4 \end{equation}
where
$$ e_1 = (0,1,1,0)~~e_2 = (0,-1,1,0)~~e_3=(-\medi,\medi,\medi,-\medi) ~~
e_4 = (1,-1,0,0). $$
Then
\begin{equation}
\xx_1 . \xx_2 = n^1_i G_{ij}n^2_j \qquad \hbox{with} \quad
 G_{ij} = {\tiny
\left( \ba{cccc}
2 & 0 & 1  & -1 \\ 0 & 2& 0 & 1 \\ 1 & 0 & 0 &0 \\ -1 & 1 & 0& 0
\ea \right)}.
\end{equation}
Let us define the cocycle
function $\epsilon (\xx,\yy) = \pm 1$ with the following properties
\begin{equation} \epsilon(\xx,\yy) = (-1)^{\xx . \yy} \epsilon(\yy,\xx)
 \label{comm} \end{equation}
\begin{equation} \epsilon(\xx ,\yy ) \epsilon(\xx +\yy , \zz ) =
\epsilon( \xx ,\yy +\zz ) \epsilon(\yy , \zz )
 \label{asociat} \end{equation}
 Choosing $\epsilon(\xx_1,\xx_2) =
(-1)^{n^1_i M_{ij} n^2_j } $ these  properties are ensured if
 $ M_{ij} - M_{ji} \equiv G_{ij} (mod~2)$. By taking
\begin{equation}
M_{ij} = {\tiny \left( \ba{cccc}
  0 & 0& 0& 0 \\ 0& 0 & 0 & 0 \\ 1& 0 &-1&1 \\
-1 & 1 &1 & 0  \ea \right) }\qquad \end{equation}
we have
\begin{equation}
\epsilon(\xx_1, \xx_2) = (-1)^{ \beta_1 (\beta_2 -m_2) +r_1 (r_2 -\alpha_2 )}.
\end{equation}
The cocycle operator is \cite{goddard}
\begin{equation} c(\xx_1) = \sum_{\xx_2} \epsilon( \xx_1 , \xx_2 ) |\xx_2
\rangle  \langle \xx_2 |
 \label{} \end{equation}
where $|\xx_2 \rangle $ is a state with eigenvalues of ghost, fermion,
matter and Liouville charges given by $\xx_2$.
The operators ${\cal O}_{\xx}$  are
redefined to $\ti{\cal O}_{\xx}={\cal O}_{\xx}c(\xx )$ so that
\begin{equation}
\ti{\cal O}_{\xx_1}\ti{\cal O}_{\xx_2}=\epsilon(\xx_1,\xx_2)
{\cal O}_{\xx_1+\xx_2} c(\xx_1+\xx_2).
\end{equation}
 This redefinition is implicitly understood when we compute the
algebra of discrete states.

\renewcommand{\arraystretch}{1}

\end{document}